\documentclass[12pt]{article}
\usepackage{amsmath}
\usepackage{amssymb}
\usepackage{dsfont}
\usepackage{cite}

\def\nn{\nonumber}

\def\a{\alpha}

\def\Lam{\Lambda}

\def\la{\lambda}
\def\om{\omega}

\def\vp{\varphi}
\def\vt{\vartheta}

\def\wh{\widehat}
\def\wt{\widetilde}
\def\ov{\overline}
\def\br{\breve}

\def\p{\partial}

\def\BC{{\mathbb C}}

\def\BR{{\mathbb R}}
\def\BN{{\mathbb N}}

\def\cla{{\mathcal A}}

\def\clc{{\mathcal C}}

\def\cll{{\mathcal L}}

\def\clq{{\mathcal Q}}
\def\clr{\mathcal{R}}

\def\im{{\rm Im\ }}
\def\spa{{\rm Span}}
\def\diag{\mathrm{diag}}
\newcommand{\E}{\mathrm{e}}
\newcommand{\I}{\mathrm{i}}

\newtheorem{Pa}{Paper}[section]
\newtheorem{Tm}[Pa]{{\bf Theorem}}

\newtheorem{Cy}[Pa]{{\bf Corollary}}
\newtheorem{Rk}[Pa]{{\bf Remark}}

\newtheorem{Ee}[Pa]{{\bf Example}}
\newtheorem{Dn}[Pa]{{\bf Definition}}

\newtheorem{Pn}[Pa]{{\bf Proposition}}

\title{GBDT with nontrivial seeds: explicit solutions   of   the focusing NLS equations and the corresponding Weyl functions}

\author{Alexander Sakhnovich \footnote{This research    was supported by the
Austrian Science Fund (FWF) grant  DOI: 10.55776/Y963.}}

\date{}
\begin{document}
\maketitle

\begin{flushright}
Faculty of Mathematics,
University
of
Vienna, \\
Oskar-Morgenstern-Platz 1, A-1090 Vienna,
Austria, \\
e-mail: {\tt oleksandr.sakhnovych@univie.ac.at}
\end{flushright}

\vspace{0.3em} 

{\it Dedicated to the memory of Rien Kaashoek and to the times of our inspiring joint research.}

\vspace{0.3em}

\begin{abstract}  Our GBDT (generalised B\"acklund-Darboux transformation) approach is used
to construct explicit solutions of the focusing nonlinear Schr\"odinger (NLS) equation in the case of the exponential seed
$a\E^{2\I(cx +dt)}$. The corresponding Baker-Akhiezer functions and evolution of the Weyl functions
are obtained as well. In particular, the solutions, which appear in the study of rogue waves,
step-like solutions and $N$-modulation solutions of the NLS equation are considered.
This work is an essential development of our joint work with Rien Kaashoek and Israel Gohberg,
where the seed was trivial, as well as several other of our previous works.
\end{abstract}

\vspace{0.2em}

Mathematics Subject Classification. Primary 35Q55; Secondary 34B20, 34L40.

\vspace{0.2em}

{\bf Keywords.} Focusing NLS, nontrivial seed, Darboux matrix, step-like solutions, $N$-modulation solutions, evolution of Weyl functions.

\section{Introduction} \label{intro}
\setcounter{equation}{0}

Focusing nonlinear Schr\"odinger (NLS) equation
\begin{align} &       \label{1.1}
\I v_t+\frac{1}{2}v_{xx}+|v|^2 v=0,
 \end{align} 
where $v_t(x,t):=\frac{\p}{\p t}v(x,t)$ and $\I$ stands for the  imaginary unit ($\I^2=-1$), is one of the most well-known integrable
equations. (Sometimes we omit the word ``focusing" further in the text and write NLS equation or NLS.)
B\"acklund-Darboux (and related) transformations  have been actively used in order to construct explicit solutions of linear and nonlinear  integrable equations including NLS. 
See, for instance, \cite{Ci, CoIv, Gek, GeT, Gu, KaG, KoSaTe, SaSaR} and various references therein for several
approaches to such transformations.  Our generalised B\"acklund-Darboux transformation (GBDT) was first introduced (for many important integrable equations) 
in \cite{SaA93, SaA2}. See also \cite{ALS1, ALS2, GKS2, KoSaTe, SaSaR} and references therein
for further developments.

The case of B\"acklund-Darboux  transformations of the trivial initial solution $v_0(x,t)\equiv 0$ (so called trivial ``seed") 
is the best studied case. In our joint works with Rien Kaashoek (and Israel Gohberg),
many of the related problems for the focusing NLS \eqref{1.1}, defocusing NLS as well as for the auxiliary Dirac systems have been studied in detail
(see, e.g.,  \cite{GKS1, GKS2, GKS3} as well as some further references in \cite{GKS3}).

Explicit solutions of direct and inverse spectral problems for auxiliary (to the focusing matrix NLS) skew-selfadjoint Dirac systems
\begin{align} &       \label{1.3}
y^{\prime}(x, z )=G(x,z)y(x,
z ),  \quad G(x,z)=\I zj + jV (x), \\ 
&       \label{1.4}
j=\begin{bmatrix}I_p & 0 \\   0 & -I_p\end{bmatrix}, \quad V(x)=\begin{bmatrix}0 & v(x) \\  v(x)^* & 0\end{bmatrix},
 \end{align} 
as well as explicit GBDT transformed solutions of matrix NLS (for the case of a trivial seed)   
were  studied in one of our first joint papers with I. Gohberg and M.A. Kaashoek \cite{GKS2}.  In the formula \eqref{1.3} above,  $y^{\prime}:=\frac{\p}{\p x}y$,  
$I_{p}$ is the $p \times p$  identity
matrix, and  $v(x)$ is a $p \times p$ matrix valued function (matrix function). The GBDT results for Dirac systems from \cite{GKS2} were further developed  
in \cite{FKRS}. According to \cite[Corollary 3.6]{FKRS}, solutions $v(x,t)$ of the focusing matrix NLS constructed in \cite{GKS2} via GBDT of the trivial solution  (so called
pseudo-exponential solutions) tend to zero when $x$ tends to infinity. These $v(x,t)$ are generated for each $t$ by the quadruples $\{\a, S(0,t), \vt_1(t), \vt_2(t)\}$
in the terminology of \cite{FKRS}. Moreover, from \cite{GKS2} or \cite{FKRS} it is easy to see that the matrix functions $v(-x,t)^*$ are generated by the quadruples 
$\{\a, S(0,t), \vt_2(t), \vt_1(t)\}$, that is, $v(-x,t)^*$ is pseudo-exponential as well. {\it Thus, we have $\lim_{|x|\to \infty} v(x,t)=0$ for the pseudo-exponential solutions.}
In \cite{BeTo}, the semiclassical scaling for the
solutions of NLS was studied under this assumption.  Evolution of the Weyl functions corresponding  to the auxiliary Dirac systems with  the GBDT-transformed
solutions $v(x,t)$ in the case  of a trivial seed follows from the results of \cite{GKS2}. Here, we obtain evolution of the Weyl functions in the case of the GBDT-transformed
solutions $v(x,t)$, where the seed is given by \eqref{1.2}.
We note that in this paper (similar to \cite{SaA2} and some other important papers on the topic) the notation $A$ is used
instead of $\a$.

Various pseudo-rational (i.e., expressed via some rational term multiplied by $\E^{\I t}$) rogue wave type solutions of NLS  have been studied using iterative Darboux transformations with the nontrivial seed $v_0=\E^{\I t}$
 in the interesting works \cite{Akh1, AnkAkh1,    GuoLing}.
 A modification 
\begin{align} &       \label{rf1}
\I \psi_t+\frac{1}{2}\psi_{xx}+(|\psi|^2-1) \psi=0
 \end{align} 
of the classical focusing NLS is often used for ``the study of spatially localized perturbations of Stokes waves" \cite{BiLM} and in order to study rogue waves, in particular.
{\it Equation \eqref{rf1} is called the NLS equation in  a ``rotating frame" in \cite{BiM} and we will use the abbreviation rfNLS for it.} Several versions of B\"acklund-Darboux transformations of the seed (background) solution $\psi_0(x,t)\equiv 1$ of the rfNLS are discussed in the papers \cite{BiLM, BiM} (see the references therein).
Clearly, if $v(x,t)$ satisfies \eqref{1.1}, then $\psi(x,t)=\E^{-\I t}v(x,t)$ satisfies rfNLS. Thus, pseudo-rational  solutions of NLS from  \cite{Akh1, AnkAkh1,   GuoLing}
turn into rational solutions of  rfNLS. As is easy to see,  one may consider the equation $ v_t+\frac{1}{2}v_{xx}+(|v|^2-2d) v=0$, where $d\in \BR$ and $\BR$ is the set of real values,
in the same way. The solutions considered in \cite{Akh1, AnkAkh1,   BiLM, BiM, GuoLing} have (after simple modification sometimes) the important property
\begin{align} &       \label{rf2}
\lim_{x \to \pm \infty}\E^{-\I t}v(x,t)=1.
 \end{align} 
 We see that the cases of solutions of \eqref{1.1}, where $\lim_{x\to \infty}\E^{-2\I d t}v(x,t)\not=0$ or $\lim_{x\to -\infty}\E^{-2\I d t}v(x,t)\not=0$ and especially the
 case
\begin{align} &       \label{rf3}
\lim_{x \to  \infty}\E^{-2\I d t}v(x,t)=\lim_{x \to - \infty}\E^{-2\I d t}v(x,t)\not=0
 \end{align}
 are of essential interest in the case of rogue waves.
 
 Another important case  of the explicit NLS  solutions recovered starting with the initial solution $\E^{\I t}$ is the case of the periodic with respect to $x$ ($N$-modulation)
 solutions (see \cite{Akh, Tr} and \cite[Appendix C]{SaA2}).
 
In this paper, we apply  our generalised B\"acklund-Darboux transformation (GBDT) to the two parameter family of initial solutions  (seeds) of NLS \eqref{1.1} of the form
\begin{align} &       \label{1.2}
v_0(x,t)=a\E^{2\I(cx +dt)}, \quad |a|^2=2(d+c^2) \quad (a\in \BC, \, a\not= 0,\, c,d\in \BR).
 \end{align} 
Here, $\BC$ is the set of complex values (i.e., complex plane). Our GBDT is an iterated
binary transformation (and so the iterations
are not required).

Skew-selfadjoint Dirac system   \eqref{1.3}, \eqref{1.4} is a particular case of AKNS or Zakharov-Shabat systems
from soliton theory and appears (itself or its subcases) as auxiliary linear system not only for the focusing NLS but also for the
mKdV and sine-Gordon equations, et cetera.
Recently, we considered (in \cite{ALS1}) GBDT for systems \eqref{1.3}, \eqref{1.4} with the seeds
$v_0(x)=a\E^{\I c x}I_p$. In the following text, we use the results on the GBDT for the focusing NLS equation from \cite{SaA2}. 
Some helpful heuristics comes from the considerations of \cite{ALS1} as well as from \cite{GKS1, GKS2}. We restrict our presentation to the scalar case
but a generalisation to the matrix NLS is straightforward.

In the next Section \ref{GBDT}, we obtain a procedure for the construction of  explicit solutions of NLS using GBDT transformations of the seeds of the form \eqref{1.2}.
In Section \ref{Ex}, we construct interesting examples and study their asymptotics. We consider examples satisfying \eqref{1.2}, step-like examples as well
as a much wider class of the $N$-modulation solutions (with an essentially wider class of seeds) than in \cite{SaA2}.
Baker-Akhiezer functions and evolution of the Weyl functions are given in Section \ref{Evol}.

The set of  positive integers is denoted by $\BN$. As usual, $\delta_{ik}$ is Kronecker delta,
the notation $\ov{a}$ means the complex conjugate of $a$ and $A^*$ means the complex conjugate transpose of the matrix $A$.
The notation $\BC_+$ stands for the upper half-plane of $\BC$: $\BC_+=\{z:\, z\in \BC, \,\, \Im(z)>0\}$, where $\Im(z)$ is the imaginary part of $z$.
By  $\BC^{n\times k}$ we denote the class of $n\times k$ matrices with complex-valued  entries and $\BC^n$ is used instead of $\BC^{n\times 1}$.
The inequality $S>0$ ($S<0$) for some matrix $S$ means that $S$ is positive (negative) definite. 
The notation $\diag\{\cla_1,\cla_2, \ldots\}$ is used for the diagonal matrix with the entries $\cla_1,\cla_2,\ldots$ on the main diagonal.

\section{GBDT for NLS with nontrivial seeds} \label{GBDT}
\setcounter{equation}{0}
{\bf 1.} NLS equation \eqref{1.1} is equivalent to the compatibility condition 
\begin{align} &       \label{2.1}
G_t(x,t,z)-F_x(x,t,z)+[G(x,t,z),F(x,t,z)]=0 
 \end{align}
$\big([G,F]:=GF-FG\big)$ of the auxiliary linear systems 
\begin{align} &       \label{2.1+}
y_x=G(x,t,z)y, \quad y_t=F(x,t,z)y,
\end{align}
where
\begin{align} &       \label{2.2}
G(x,t,z)=\I zj + jV (x,t), \quad
j=\begin{bmatrix}1 & 0 \\   0 & - 1\end{bmatrix}, \quad V(x,t)=\begin{bmatrix}0 & v(x,t) \\  \ov{v(x,t)} & 0\end{bmatrix},
\\ &       \label{2.3}
F(x,t,z)=\I\big(-z^2j+\I z j V(x,t)+\big(V_x(x,t)+jV(x,t)^2\big)\big/2\big).
 \end{align}
 Equality \eqref{2.1} is the well-known {\it zero curvature equation}, where the choice of $F$ and $G$ for NLS is not unique.
 Our  matrix functions $G$ and $F$ are defined by $G(x,t,z)=J\br G(x,t,-z)J$ and $F(x,t,z)=J\br F(x,t,-z)J$,
 where $\br G$ and $\br F$ are the corresponding (to NLS equation) $G$ and $F$ in the notations of \cite{SaA2}, 
\begin{align} &       \label{J} 
 J=\begin{bmatrix}0 & 1 \\   1 & 0\end{bmatrix},
\end{align} 
 and $u$ from  \cite{SaA2} is substituted by $v$ here.

GBDT  is determined by the initial solution $v_0(x,t)$  of the nonlinear equation \eqref{1.1} or, equivalently, by 
\begin{align}& \label{2.4-}
V_0(x,t)=\begin{bmatrix}0 & v_0(x,t) \\  \ov{v_0(x,t)} & 0\end{bmatrix},
\end{align}
and by a triple of parameter matrices $ \{A, S(0,0), \Lam(0,0)\}$ (see, e.g., \cite{SaA2} or \cite[Subsection 1.1.3]{SaSaR}).
Here, $A, S(0,0)\in \BC^{n\times n}$  $(S(0,0)=S(0,0)^*, \,\, n\in \BN)$, $\Lam(0,0)\in \BC^{n\times 2}$ 
and the matrix identity
\begin{align}& \label{2.4}
AS(0,0)-S(0,0)A^*=\I \Lam(0,0) \Lam(0,0)^*
\end{align}
holds. The matrix $A$ is a so called {\it generalised matrix eigenvalue}, and $\break \Lam(x,t)\in \BC^{n\times 2}$ is determined
by its value $\Lam(0,0)$ and by the linear systems:
\begin{align} \label{2.5}
\Lam^{\prime}(x,t)=&-\I A \Lam(x,t)j-\Lam(x,t)j V_0(x,t) \quad (\Lam^{\prime}:=\frac{\p}{\p x}\Lam),
\\  \nn
\dot{\Lam}(x,t)=&\I A^2 \Lam(x,t)j+A\Lam(x,t)jV_0(x,t)
\\ &\label{2.6}
-\frac{\I}{2}\Lam(x,t)V_0^{\prime}(x,t)-\frac{\I}{2}\Lam(x,t)jV_0(x,t)^2 \quad (\dot{\Lam}:=\frac{\p}{\p t}\Lam).
\end{align}
Systems \eqref{2.5} and \eqref{2.6} follow from the equations \cite[(10) and (12)]{SaA2} for $\Pi=\Lam J$,
where $J$ is given by \eqref{J}.
Now, the matrix function $S(x,t)$ is determined (see \cite[p. 701]{SaA2}) by its value $S(0,0)$ and by the derivatives
\begin{align}& \label{2.7}
S^{\prime}(x,t)=\Lam(x,t)j\Lam(x,t)^*, \\
&  \label{2.8}
 \dot{S}(x,t)=-A\Lam(x,t)j\Lam(x,t)^*-\Lam(x,t)j\Lam(x,t)^*A^*+\I \Lam(x,t)jV_0(x,t)\Lam(x,t)^*.
\end{align}
Relations \eqref{2.5}--\eqref{2.8} imply the identity 
\begin{align}& \label{2.9}
AS(x,t)-S(x,t)A^*=\I \Lam(x,t) \Lam(x,t)^*
\end{align}
at each $x$ and $t$ (and not only at $x=0$, $t=0$ as in \eqref{2.4}).  We have also $S(x,t)=S(x,t)^*$.

Let $v_0(x,t)$ satisfy \eqref{1.1} and assume that relations \eqref{2.4}--\eqref{2.8} are valid.
Denote the columns of $\Lam$ by $\Lam_k$: $\,\Lam(x,t)=\begin{bmatrix}\Lam_1(x,t)  & \Lam_2(x,t)\end{bmatrix}$.
Then (see \cite[Theorem 1]{SaA2}), the function 
\begin{align}& \label{2.10}
v(x,t)=v_0(x,t)+2 \Lam_1(x,t)^* S(x,t)^{-1}\Lam_2(x,t)
\end{align}
satisfies (in the points of invertibility of $S(x,t)$) the focusing NLS equation as well. It is called {\it a GBDT-transformed solution of} \eqref{1.1}. 
We also say that $v(x,t)$ is generated by the triple $ \{A, S(0,0), \Lam(0,0)\}$ and the seed $v_0(x,t)$.
\begin{Rk}\label{RkInv}  If $S(0,T)>0$ for some $T\in \BR$, then $($according to \cite[Proposition 2.4]{ALS1}$)$ $S(x,T)$ is invertible for all $x\in \BR$.
In a similar way, using \eqref{2.7} and \eqref{2.9}, it is proved that $S(x,T)$ is invertible for all $x\in \BR$ if $S(u,T)>0$ for some $u\in \BR$ or
 if $S(u,T)<0$ for some $u\in \BR$. Moreover, $S(x,T)>0$ if $S(u,T)>0$ and  $S(x,T)<0$ if $S(u,T)<0$.

The invertibility of $S(x,t)$ for all $x,t\in \BR$ is discussed in Proposition \ref{PnInv}.
\end{Rk}
We present the so called  Darboux matrix (i.e., Darboux matrix function) as a transfer matrix function in Lev Sakhnovich \cite{SaL1, SaL2} form:
\begin{align}& \label{2.9+}
w_A(x,t,z)=I_2-\I \Lam(x,t)^*S(x,t)^{-1}(A-z I_n)^{-1}\Lam(x,t).
\end{align}
Here, $v_0(x,t)=\wh u(x,t)$,
$\, \Lam(x,t)=\wh \Pi(x,t) J$,  $S(x,t)=\wh S(x,t)$, and $w_A(x,t,z)=J\wh w_A(x,t,z)J$,
 where $\wh u(x,t)$, $\,\wh \Pi(x,t), \, \wh S(x,t)$, and $\wh w_A(x,t,z)$ are $u(x,t)$, $\, \Pi(x,t)$, $\, S(x,t)$, and $w_A(x,t,z)$ in the notations of Example 1 from 
 \cite[pp. 297-300]{ALS1+}.
According to \cite[p. 298]{ALS1+}, our $G(x,t,z)$  and $F(x,t,z)$ given by \eqref{2.2} and \eqref{2.3} coincide with $J\wt G(x,t,z)J$ and $J \wt F(x,t,z)J$  for
$\wt G(x,t,z)$ and $\wt F(x,t,z)$ from \cite{ALS1+} (after substitution of $v$ instead 
of $\wt u$ considered in \cite{ALS1+}). Now, formula (1.34) (see also Theorem 3.2, both from \cite{ALS1+}) yields
\begin{align}& \label{2.9++}
\frac{\p}{\p x}w_A(x,t,z)=G(x,t,z)w_A(x,t,z)-w_A(x,t,z)G_0(x,t,z), \\
& \label{2.9!}
 \frac{\p}{\p t}w_A(x,t,z)=F(x,t,z)w_A(x,t,z)-w_A(x,t,z)F_0(x,t,z),
\end{align}
where $G_0$ and $F_0$ are given by \eqref{2.2} and \eqref{2.3}, respectively, after the substitution of $V_0$ instead of $V$ there.
Relations \eqref{2.9++} and \eqref{2.9!} are characteristic for Darboux matrices. If the $2\times 2$ matrix function $w_0(x,t,z)$
satisfies
\begin{align}& \label{2.10!}
\frac{\p}{\p x}w_0(x,t,z)=G_0(x,t,z)w_0(x,t,z), \quad \frac{\p}{\p t}w_0(x,t,z)=F_0(x,t,z)w_0(x,t,z),
\end{align}
then the product
\begin{align}& \label{2.10--}
w(x,t,z)=w_A(x,t,z)w_0(x,t,z)
\end{align}
is the Baker-Akhiezer or wave function of the GBDT-transformed system and satisfies relations
\begin{align}& \label{2.10-}
\frac{\p}{\p x}w(x,t,z)=G(x,t,z)w(x,t,z), \quad \frac{\p}{\p t}w(x,t,z)=F(x,t,z)w(x,t,z).
\end{align}
\begin{Rk}\label{Rk2.2}   Recall that our GBDT formulas  were first introduced $($for many important integrable equations including NLS$)$ 
in \cite{SaA93, SaA2}.   Several years later similar formulas for the solutions of NLS $($matrix identities of the \eqref{2.9} type
and relations of the \eqref{2.10} type, in particular$)$  appeared in the work \cite{Man} by M.~Ma\~nas, see  some discussion
in the Subsection 7.2 $($dedicated to GBDT$)$ of the important survey \cite{Ci}.  Much later $($starting ca. from the year 2010,
see, e.g., \cite{ChMH, DiMH1}$)$ similar  formulas appeared in various papers by F.  M\"uller-Hoissen and coauthors.
However, neither Darboux matrices nor Baker-Akhiezer matrix functions are constructed in \cite{ChMH, DiMH1, Man}.
Note
that ``wavefunctions"  in \cite{Man} are not the classical wave functions, do not depend on the spectral parameter $z$
and cannot be used  for our purposes here.

It should be said that Darboux matrices and  Baker-Akhiezer matrix functions play a fundamental role in GBDT.
For instance, the study of the interrelations between Weyl functions and nonlinear equations in our joint works \cite{GKS1, GKS2}
with
Rien Kaashoek and Israel Gohberg
and in the present paper is based on those constructions. The present work is an interesting development of \cite{GKS1, GKS2}.  Here, Weyl functions and their evolution are constructed for the case
of the solutions of NLS obtained via GBDT of nontrivial seeds $($including  $N$-modulated solutions and rogue waves$)$
instead of the case of trivial seeds in \cite{GKS1, GKS2}. 

\end{Rk}

{\bf 2.} A pair of the matrices $\{A,\Lam\}$ is called controllable or full range if $\break \spa \bigcup_{k=0}^{n-1}\im A^k\Lam=\BC^n,$
where $\im$ means image and $A$ and $\Lam$ are  $n\times n$ and $n\times m$, respectively, matrices. (Note that $m=2$
in our considerations.)
\begin{Pn}\label{PnInv} Let $A$, $\Lam(x,t)$ and $S(x,t)$ satisfy \eqref{2.4}--\eqref{2.6} and assume that the 
pair $\{A,\Lam(x_0, t_0)\}$ is controllable for some fixed values $x_0$ and $t_0$.  Then, $S(x_0,t_0)$ is invertible.
\end{Pn}
{\it Proof.} Assume that $S(x_0,t_0)$ is not invertible. Then, we have $S(x_0,t_0)f=0$ for some $f\in \BC^n$, $f\not=0$.
Now,  \eqref{2.9} yields $f^*\Lam(x_0, t_0)\Lam(x_0, t_0)^*f=~0$, that is, $\Lam(x_0, t_0)^*f=0$. Relations $S(x_0,t_0)f=0$,
$\Lam(x_0, t_0)^*f=0$ and \eqref{2.9} imply in turn that $S(x_0,t_0)A^*f=0$. Hence $\Lam(x_0, t_0)^*A^*f=0$. By induction,
we obtain $\Lam(x_0, t_0)^*(A^*)^kf=0$ and $f^*A^k\Lam(x_0, t_0)=0$ for $k\geq 0$. In view of the controllability
of $\{A,\Lam(x_0, t_0)\}$, it follows that $f=0$ and  we come to a contradiction. $\Box$

The proof of the proposition above gives a stronger corollary in the case of the diagonal matrices $A=\diag\{\cla_1,\cla_2, \ldots, \cla_n\}$.
By $\cll_k$, we denote below the $k$-th row of  $\Lam$: $\Lam=\{\cll_k\}_{k=1}^n$.
\begin{Cy}\label{CyDiag}  Let $A=\diag\{\cla_1,\cla_2, \ldots, \cla_n\}$, $\Lam(x,t)$ and $S(x,t)$ satisfy \eqref{2.4}--\eqref{2.6} and assume that
\begin{equation}\label{2.27--}
 \cla_i\not= \cla_k \quad (i\not=k), \quad \cla_i\not=\ov{\cla_k},  \quad 
\cll_k(0,0)\not=0 \quad  (1\leq i,k \leq n).
\end{equation}
Then, for $S(x,t)=\{s_{ik}(x,t)\}_{i,k=1}^n$ and its entries $s_{ik}$ we have
\begin{align}\label{2.27-}&
s_{ik}(x,t)=\frac{\I}{\cla_i - \ov{\cla_k}}\cll_i(x,t)\cll_k(x,t)^*, \quad \det S(x,t)\not=0.
\end{align}
\end{Cy}
{\it Proof}. The  equality in \eqref{2.27-} is immediate from the identity  \eqref{2.9}.

If the inequality in \eqref{2.27-} does not hold for some $x,t\in \BR$, we have $S(x,t)f=0$
for some $f=\{f_k\}_{k=1}^n\not=0$. Then, similar to the proof of Proposition \ref{PnInv} we obtain
\begin{align}\label{p1}&
f^*A^i\Lam(x, t)=0 \quad {\mathrm{for}}\quad  i\geq 0.
\end{align}
It follows from \eqref{p1} that
\begin{align}\nn
\sum_{i=1}^n\ov{\clc_i}  f^*A^i\Lam(x, t)&=\sum_{i=1}^n \ov{\clc_i}\sum_{k=1}^n\ov{f_k}\, \cla_k^i\cll_k(x,t)
\\ \label{p2}&
=\clc^* \{\cla_k^i\}_{i,k=1}^n\{\ov{f_k}\cll_k(x,t)\}_{k=1}^n=0
\end{align}
for any $\clc=\{\clc_i\}_{i=1}^n\in \BC^n$. Here, $\{\ov{f_k}\cll_k(x,t)\}_{k=1}^n\in \BC^{n \times 2}$.
We rewrite \eqref{p2} in the form
\begin{equation}\label{p3}
\clc^* \{\cla_i^k\}_{i,k=1}^n g=0, \quad g:=\diag\{\ov{f_1}, \ov{f_2},\ldots ,\ov{f_n}\}(\a_1 \Lam_1(x,t)+\a_2\Lam_2(x,t)).
\end{equation}
Clearly, \eqref{p3} holds for any $\a_1,\a_2 \in \BC$.

Next, we note that equations \eqref{2.5} and \eqref{2.6} yield
\begin{align}\label{p4}&
\cll_k^{\prime}(x,t)=-\cll_k(x,t)\big(\I\cla_k j+j V_0(x,t)\big), \\
\label{p5}&
 \dot{\cll_k}(x,t)=\cll_k(x,t)\big(\I\cla_k^2 j+\cla_k j V_0(x,t)-\I (V_0^{\prime}(x,t)+jV_0(x,t)^2)/2\big).
\end{align}
Hence, the inequality $\cll_k(x,t)\not=0$ follows from the inequality $\cll_k(0,0)\not=0$ in \eqref{2.27--}.

Finally, recall that \eqref{p3} holds for any $\clc\in \BC^n$ and $\a_1,\a_2\in \BC$.
In view of the first inequality in \eqref{2.27--} the matrix $ \{\cla_i^k\}_{i,k=1}^n$ is non-degenerate
and so \eqref{p3} yields $g=0$. Since $\cll_k(x,t)\not=0$ $(1\leq k \leq n)$, $\a_1$ and $\a_2$
may be chosen so that all the entries of $\a_1 \Lam_1(x,t)+\a_2\Lam_2(x,t)$ are nonzero
at our fixed values of $x$ and $t$. Then, taking into account the definition of $g$ in \eqref{p3}
and the equality $g=0$, we obtain $f=0$. The inequality $\det S(x,t)\not= 0$ is proved by negation. $\Box$

{\bf 3.} Next, we consider the case of $v_0(x,t)$ of the form \eqref{1.2} and derive explicit expressions for $\Lam_1(x,t)$ and $\Lam_2(x,t)$.
System \eqref{2.5} may be rewritten in the form
\begin{align}& \label{2.10+}
\Lam_1^{\prime}=-\I A\Lam_1+\ov{v_0}\Lam_2, \quad \Lam_2^{\prime}=\I A\Lam_2-{v_0}\Lam_1.
\end{align}
The exponents $\E^{\pm \I x Q}$, where
\begin{align}& \label{2.11}
 Q^2=(A-cI_n)^2+|a|^2I_n , \quad AQ=QA.
\end{align}
play an essential role in the formulas for $\Lam_1(x,t)$ and $\Lam_2(x,t)$ in \cite{ALS1}.
The proposition below follows from \cite[Proposition 3.3]{ALS1}
\begin{Pn}\label{PnQ}
Let $a\not= 0$, $c\in \BR$, an $n\times n$ matrix $A$ be given and the inequality $\det\big((A-cI_n)^2+|a|^2I_n\big)\not=0$
be valid.
Then,  there is an $n\times n$ matrix $Q$ such that \eqref{2.11} holds.
\end{Pn}
 It follows from \eqref{1.2} and relation $V_0=\begin{bmatrix}0 & v_0 \\  v_0^* & 0\end{bmatrix}$
that
\begin{align}& \label{2.12}
jV_0(x,t)^2=|a|^2j, \quad V_0^{\prime}(x,t)=2\I c j V_0(x,t).
\end{align}
Thus, the dependence \eqref{2.6} of $\Lam$ on $t$  may be rewritten in the form
\begin{align}& \label{2.13}
\dot{\Lam}(x,t)=\I \Big(A^2-\frac{|a|^2}{2}I_n\Big)\Lam(x,t)j+(A+cI_n)\Lam(x,t)jV_0(x,t).
\end{align}
Using the equality $|a|^2=2(d+c^2)$ (see \eqref{1.2}), we rewrite \eqref{2.13} as
\begin{align}& \label{2.14}
\dot{\Lam}_1(x,t)=\I \big(A^2-(d+c^2)I_n\big)\Lam_1(x,t)-\ov{v_0(x,t)}(A+cI_n)\Lam_2(x,t),
\\ & \label{2.15}
\dot{\Lam}_2(x,t)=-\I \big(A^2-(d+c^2)I_n\big)\Lam_2(x,t)+{v_0(x,t)}(A+cI_n)\Lam_1(x,t).
\end{align}
Hence, setting $\Psi_1(x,t)=\E^{\I d t}\Lam_1(x,t)$ and $\Psi_2(x,t)=\E^{-\I d t}\Lam_2(x,t)$ and using \eqref{2.11}, we derive
\begin{align}& \label{2.16}
\ddot{\Psi}_k(x,t)=-(A+c I_n)^2Q^2\Psi_k \quad (k=1,2),
\end{align}
that is, the expressions $\exp\{\pm \I t(A+cI_n)Q\}$ should appear in the formulas for $\Lam_k(x,t)$. Indeed,
we will show that the vectors
\begin{align}& \label{2.17}
{\Lam}_1(x,t)=\E^{-\I(cx+dt)}\big(\E^{\I\clr(x,t)}h_1+\E^{-\I\clr(x,t)}h_2\big),
\\ & \label{2.18}
{\Lam}_2(x,t)=\E^{\I(cx+dt)}\big(\E^{\I\clr(x,t)}h_3+\E^{-\I\clr(x,t)}h_4\big),
\end{align}
where
\begin{align}& \label{2.19}
\clr(x,t)=(xI_n-t(A+cI_n))Q, \quad h_k\in \BC^n \quad (1\leq k\leq 4),
\\ & \label{2.20}
h_3=(\I/\ov{a})(Q+A-c I_n)h_1, \quad h_4=(\I/\ov{a})(A-Q-c I_n)h_2,
\end{align}
satisfy \eqref{2.14} and \eqref{2.15}. The commutativity property $AQ=QA$, which could be omitted in
the considerations of \cite{ALS1}, is essential here.
\begin{Tm}\label{ExplLam} Let $v_0$ be given by \eqref{1.2} and let $\Lam_1$ and $\Lam_2$ be given
by \eqref{2.17} and \eqref{2.18}, where the relations \eqref{2.11}, \eqref{2.19} and \eqref{2.20} hold.
Then, $\Lam_1$ and $\Lam_2$ satisfy linear systems \eqref{2.10+} and \eqref{2.14}, \eqref{2.15} or, equivalently,
systems \eqref{2.5} and \eqref{2.6}.
\end{Tm}
{\it Proof}. For $\Lam_1$ and $\Lam_2$ given by  \eqref{2.17} and \eqref{2.18}, where \eqref{2.19} and \eqref{2.20}
holds, we have
\begin{align}\nn
{\Lam}_1^{\prime}(x,t)=&-\I A {\Lam}_1(x,t) +\I \E^{-\I(cx+dt)}\big(\E^{\I\clr(x,t)}(A-c I_n+Q)h_1
\\ & \nn
+\E^{-\I\clr(x,t)}(A-c I_n-Q)h_2\big)=-\I A {\Lam}_1(x,t) +(\I/\ov{a})\ov{v_0(x,t)}\E^{\I(cx+dt)}
\\ \nn &
\times 
\big(\E^{\I\clr(x,t)}(A-c I_n+Q)h_1
+\E^{-\I\clr(x,t)}(A-c I_n-Q)h_2\big)
\\  \label{2.21}
=& -\I A {\Lam}_1(x,t)+\ov{v_0(x,t)}\Lam_2(x,t),
\end{align}
that is, the first equation in \eqref{2.10+} is valid. 

In order to prove the second equality   in \eqref{2.10+}, we note that
formula \eqref{2.11} implies the relation
\begin{align}& \label{2.23}
(\I/\ov{a})(Q+A-c I_n)(Q-A+cI_n)=(\I/\ov{a})(Q^2-(A-c I_n)^2)=\I a.
\end{align}
Hence, in view of \eqref{2.20} we obtain
\begin{align}& \label{2.24}
(\I/{a})(-A+c I_n+Q)h_3=-h_1, \quad (\I/{a})(-A+c I_n-Q)h_4=-h_2.
\end{align}
Using \eqref{2.24} as well as \eqref{1.2} and \eqref{2.17}--\eqref{2.20}, we derive
\begin{align}\nn
{\Lam}_2^{\prime}(x,t)=&\I A {\Lam}_2(x,t) +\I \E^{\I(cx+dt)}\big(\E^{\I\clr(x,t)}(-A+c I_n+Q)h_3
\\ & \nn
+\E^{-\I\clr(x,t)}(-A+c I_n-Q)h_4\big)=\I A {\Lam}_2(x,t) +(\I/{a})v_0(x,t)\E^{-\I(cx+dt)}
\\ \nn &
\times 
\big(\E^{\I\clr(x,t)}(-A+c I_n+Q)h_3
+\E^{-\I\clr(x,t)}(-A+c I_n-Q)h_4\big)
\\  \label{2.25}
=& \I A {\Lam}_2(x,t)-{v_0(x,t)}\Lam_1(x,t).
\end{align}
Thus, the second equality   in \eqref{2.10+} also holds, that is, our $\Lam_1$ and $\Lam_2$ satisfy \eqref{2.10+}.

In a similar way, we show that relations \eqref{1.2}, \eqref{2.17}--\eqref{2.20} and \eqref{2.24} yield \eqref{2.14} and
\eqref{2.15}. Indeed, for $\Lam_1$ given by \eqref{2.17} we have
\begin{align}\nn
\dot{\Lam}_1(x,t)=& \I \big(A^2-(d+c^2)I_n\big)\Lam_1(x,t)+\I \E^{-\I(cx+dt)}
\\ \nn & \times
\big(\E^{\I\clr(x,t)}(-A^2+(d+c^2)I_n-d I_n-(A+c I_n)Q)h_1
\\ & \nn
+\E^{-\I\clr(x,t)}(-A^2+(d+c^2)I_n-d I_n+(A+c I_n)Q)h_2\big)
\\ \nn =&\I \big(A^2-(d+c^2)I_n\big)\Lam_1(x,t)-\I (-\I \ov{a})(A+c I_n)\E^{-\I(cx+dt)}
\\ \nn & \times \big(\E^{\I\clr(x,t)}h_3+\E^{-\I\clr(x,t)}h_4\big)
\\ \label{2.26} =&
\I \big(A^2-(d+c^2)I_n\big)\Lam_1(x,t)-\ov{v_0(x,t)}(A+cI_n)\Lam_2(x,t),
\end{align}
that is, \eqref{2.14} is valid. For $\Lam_2$ given by \eqref{2.18} we have
\begin{align}\nn
\dot{\Lam}_2(x,t)=&- \I \big(A^2-(d+c^2)I_n\big)\Lam_2(x,t)+\I \E^{\I(cx+dt)}
\\ \nn & \times
\big(\E^{\I\clr(x,t)}(A^2-(d+c^2)I_n+d I_n-(A+c I_n)Q)h_3
\\ & \nn
+\E^{-\I\clr(x,t)}(A^2-(d+c^2)I_n+d I_n+(A+c I_n)Q)h_4\big)
\\ \nn =&-\I \big(A^2-(d+c^2)I_n\big)\Lam_2(x,t)+\I (-\I {a})(A+c I_n)\E^{\I(cx+dt)}
\\ \nn & \times \big(\E^{\I\clr(x,t)}h_1+\E^{-\I\clr(x,t)}h_2\big)
\\ \label{2.26+} =&
-\I \big(A^2-(d+c^2)I_n\big)\Lam_2(x,t)+{v_0(x,t)}(A+cI_n)\Lam_1(x,t),
\end{align}
that is, \eqref{2.15} is valid. $\Box$
\begin{Rk}\label{RkExpl}
Explicit expressions \eqref{2.17}--\eqref{2.20} for $\Lam_1$ and $\Lam_2$ together with formulas \eqref{2.7}, \eqref{2.8}
and \eqref{2.10}, where seeds $v_0(x,t)$ have the form \eqref{1.2},  provide explicit GBDT-transformed solutions $v(x,t)$ of the NLS equation
\eqref{1.1}. Moreover, using formulas \eqref{2.9+}, \eqref{2.10--} and \eqref{4.1} we obtain explicit expressions for the corresponding Baker-Akhiezer functions $w(x,t,z)$.
\end{Rk}
\begin{Rk}\label{Rk0} If $Q$ is invertible, then $h_1$ and $h_2$ are easily expressed via $\Lam(0,0)$.
Indeed, relations \eqref{2.17}-- \eqref{2.20}  imply 
$$\Lam_1(0,0)=h_1+h_2, \quad \Lam_2(0,0)=h_3+h_4=\frac{\I}{\ov{a}}\Big((A-cI_n)(h_1+h_2)+Q(h_1-h_2)\Big).$$
Hence, we have
\begin{align}\label{2.26++}&
h_1+h_2=\Lam_1(0,0), \quad h_1-h_2=-Q^{-1}\big((A-cI_n)\Lam_1(0,0)+\I \ov{a}\Lam_2(0,0)\big).
\end{align}
The expressions for $h_1$ and $h_2$ easily follow. \end{Rk}

\section{Examples} \label{Ex}
\setcounter{equation}{0}
Let us consider several examples corresponding to the case $n=1$, where the asymptotical behaviour of $v(x,t)$
greatly varies.
\begin{Ee}\label{Ee1} Assume that $n=1$,
\begin{align}\label{2.27}&
a=\I r, \,\, c=0, \,\, d=r^2/2,  \,\,  A=\I \la, \,\, h_1=g, \,\, h_2=1 \quad (r,\, \la,\, \in \BR), \\
\label{2.28} &
r\not=0,   \quad \la > |r|, \quad Q=\I \mu, \quad  \mu=\sqrt{\la^2-r^2},
\end{align}
where any of the two branches  of the square root in \eqref{2.28} may be fixed.
\end{Ee}
It follows from \eqref{2.17}--\eqref{2.20} that
\begin{align}\label{2.29}&
\Lam_1(x,t)=\E^{-\I d t}\big(g \exp\{(\I \la t-x)\mu\}+\exp\{(x-\I\la t)\mu\}\big),\\
\label{2.30} &
\Lam_2(x,t)=(1/\I r)\E^{\I d t}\big((\la +\mu)g \exp\{(\I \la t-x)\mu\}+(\la - \mu)\exp\{(x-\I\la t)\mu\}\big).
\end{align}
In view of \eqref{2.9}, we obtain
\begin{align}& \label{2.31}
S(x,t)=\frac{1}{2\la}\big(| \Lam_1(x,t)|^2+| \Lam_2(x,t)|^2\big)>0.
\end{align}
Finally, formulas \eqref{2.10} and \eqref{2.29}--\eqref{2.31} yield
\begin{align}& \label{2.32}
v(x,t)=\E^{2\I d t}\left(a+\frac{4\la\om_1(x,t)}{\I r \om_2(x,t)}\right),
\end{align}
where $d=r^2/2$ and
\begin{align}\nn 
\om_1(x,t)=&(\la+\mu)|g|^2\E^{-2x\mu}+(\la-\mu)\E^{2x\mu}+(\la-\mu)\ov{g}\E^{-2\I \la t\mu}+(\la+\mu){g}\E^{2\I \la t\mu},\\
\nn 
\om_2(x,t)=&\big((1+((\la+\mu)/r)^2\big)|g|^2\E^{-2x\mu}+\big(1+((\la-\mu)/r)^2\big)\E^{2x\mu}\\
\label{2.33} &
+2\ov{g}\E^{-2\I \la t\mu}+2{g}\E^{2\I \la t\mu}.
\end{align}
The family of the solutions $v(x,t)$ given by \eqref{2.32} is determined by the parameters $a$ (or $r$) and $\la$.
In order to derive \eqref{2.33}, we used  the equality $\la^2-\mu^2=r^2$, which also yields
\begin{align}& \label{2.33+}
(\la-\mu)^2+r^2=2\la(\la-\mu), \quad (\la+\mu)^2+r^2=2\la(\la+\mu).
\end{align}
Let us  set $\psi(x,t)$ and fix the branch of $\mu$ via the formulas:
\begin{align}& \label{2.34}
\psi(x,t)=e^{-2\I dt}v(x,t), \quad \mu>0.
\end{align}
Then, the equality $a=\I r$ and  relations \eqref{2.32}--\eqref{2.34} yield the following corollary.
\begin{Cy}\label{CyEx1} Let \eqref{2.34} hold. Then, for the NLS solutions $v(x,t)$  in Example~\ref{Ee1} we have
\begin{align}\label{2.35}&
\lim_{x\to \infty}\psi(x,t)=\lim_{x\to -\infty}\psi(x,t)=-\I r.
\end{align}
For the subcase $g=0$ $($in \eqref{2.27}$)$, we obtain $\psi(x,t)\equiv -\I r$.
\end{Cy}
We note that the functions $\psi(x,t)$ above are periodic with respect to $t$.
\begin{Rk} \label{RkPseudo} The pseudo-rational solutions $($i.e., the solutions expressed via  the seed term $a\E^{2\I(cx +dt)}$ and rational terms$)$ 
may appear in the case of the polynomial expressions $\E^{\pm \I \clr(x,t)}$, which, according
to \eqref{2.19}, happens if the matrix $Q$ is nilpotent. However, in many cases one gets $\det S(x,t)\equiv 0$ or the evident case $v(x,t) \equiv -v_0(x,t)$.
For instance, in the case of $n=3$, a $3\times 3$ matrix $Q=\{ \delta_{i+1,k}\}_{i,k=1}^3$ and $A=(c+\I |a|)I_3+\frac{1}{2\I |a|}Q^2$, we have $\det S(x,t)\equiv 0$
if
\begin{align}& \label{3.1}
h_1=\begin{bmatrix} 1 \\ 0 \\ 1\end{bmatrix}, \quad h_2= \begin{bmatrix} 0 \\ 1 \\ -1\end{bmatrix},
\end{align}
and we have $\det S(x,t)\equiv \frac{1}{64 |a|^9}$, $v(x,t) \equiv -v_0(x,t)$ if
\begin{align}& \label{3.2}
h_1=\begin{bmatrix} 0 \\ 0 \\ 1\end{bmatrix}, \quad h_2= \begin{bmatrix} 0 \\ g \\ 0\end{bmatrix} \quad (g\in \BC).
\end{align}
The case of pseudo-rational solutions will be treated in greater detail in the future work.
\end{Rk}
Example \eqref{2.27}, \eqref{2.28} for the special  case $t=0$ and $g\in \BR$ was considered in \cite[Example 4.3]{ALS1},
although we omitted there the fact that the limits of $v(x)$ at $+\infty$ and $-\infty$ coincide.
 The next two examples
were not studied earlier even for $t=0$. 
\begin{Ee}\label{Ee2} 
Let us fix $r$, which does not essentially effect the behaviour of the corresponding solution, and set
\begin{align}\label{3.3}&
a=\I, \,\, c=0, \,\, d=1/2,  \,\,  A=\I \la \,\, (0<\la<1), \,\, h_1=g, \,\, h_2=1, 
\\ & \label{3.4}
Q=\mu=\sqrt{1-\la^2}. 
\end{align}
where any of the two branches  of the square root in \eqref{3.4} may be fixed.
\end{Ee}
It follows from \eqref{1.2} and \eqref{3.3} that $v_0(x,t)=\I \E^{\I t}$.
Similar to the Example~\ref{Ee1}, the equality \eqref{2.31} holds. According to \eqref{2.20}, we have
\begin{align}\label{3.5}
h_3=-g(\I \la+\mu), \quad h_4=-(\I \la-\mu).
\end{align}
By virtue of \eqref{2.17}, we obtain
\begin{align}\label{3.6}
\Lam_1(x,t)=\E^{-\I t/2}\big(g\E^{\I \mu (x-\I \la t)}+\E^{-\I \mu (x-\I \la t)}\big).
\end{align}
Formulas \eqref{2.18} and \eqref{3.5} yield
\begin{align}\label{3.7}
\Lam_2(x,t)=-\E^{\I t/2}\big(g(\I \la +\mu)\E^{\I \mu (x-\I \la t)}+(\I \la-\mu)\E^{-\I \mu (x-\I \la t)}\big).
\end{align}
Formulas \eqref{3.3} and \eqref{3.4} imply that
\begin{align}&\nn
(\I\la +\mu)(\I \la -\mu)=-(\la^2+\mu^2)=-1,
 \\ & \nn
 1-(\I \la +\mu)^2=2\la (\la -\I \mu),\,\, 1-(\I \la -\mu)^2=2\la (\la +\I \mu).
\end{align}
Hence, using \eqref{2.31}, \eqref{3.6} and \eqref{3.7} we derive
\begin{align}\label{3.8}
S(x,t)=\big(|g|^2\E^{2\mu \la t}+\E^{-2\mu \la t}+g\la (\la -\I \mu)\E^{2\I \mu x}+\ov{g}\la (\la +\I \mu)\E^{-2\I \mu x}\big)\big/ \la .
\end{align}
Using again \eqref{3.6} and \eqref{3.7} and setting $\om(x,t)=\E^{-\I t}\Lam_1(x,t)^*\Lam_2(x,t)$, we also derive
\begin{align}\nn\om(x,t)=-\big(&|g|^2(\I \la +\mu)\E^{2\mu \la t}+(\I \la -\mu)\E^{-2\mu \la t}+g(\I\la + \mu)\E^{2\I \mu x}
\\ \label{3.9} &
+\ov{g} (\I\la - \mu)\E^{-2\I \mu x}\big).
\end{align}
Finally, formula \eqref{2.10} takes the form
\begin{align}\label{3.10}&
v(x,t)=\E^{\I t}\big(\I+2\om(x,t)\big/ S(x,t)\big).
\end{align}
\begin{Rk}\label{RkNL}
The considered case $0<\la <1$ differs essentially from the case $\la >r=1$ $($see Example \ref{Ee1}$)$.
In view of \eqref{3.8}--\eqref{3.10}, the function $\psi(x,t)=\E^{-\I t}v(x,t)$ $($as well as the function $v(x,t))$ is periodic with respect to $x$ and therefore
does not have any limits when $x$ tends to $+\infty$ or $-\infty$. $($Even $|\psi(x,t)|$ does not have any limits.$)$ The periodic in $x$ solutions
of NLS given by  \eqref{3.10} belong to the class of modulation solutions \cite{Akh, SaA2, Tr}.
\end{Rk}
\begin{Rk}\label{RkB} In the case $a=A=\I$ $(r=\la=1)$, which is intermediate between Examples \ref{Ee1} and \ref{Ee2}, we have
$($similar to some cases in Remark~\ref{RkPseudo}$)$ $v(x,t)=-v_0(x,t)=-\I \E^{\I t}$.
\end{Rk}
\begin{Ee}\label{Ee3} 
Let us   consider the case $c\not=0$ and set:
\begin{equation}\label{3.11}
a=\I, \,\,  d=1/2-c^2,  \,\,  A=2 \I,  \,\, h_1=g, \,\, h_2=1, \,\, Q=\mu=\sqrt{c^2-4\I c-3},
\end{equation}
where $\mu \in \BC$ and any of the two branches  of the square root in \eqref{3.11} may be fixed.
Clearly, conditions \eqref{3.11}  satisfy \eqref{2.11} and the second equality in \eqref{1.2}.
\end{Ee}
It follows from \eqref{3.11} and \eqref{1.2}, \eqref{2.19}, \eqref{2.20} that 
\begin{align}\nn &
v_0(x,t)=\I \E^{2\I(cx+dt) }=\I \exp\{\I(2cx-2c^2t+t)\}, \quad \I\clr(x,t)=\I \mu\big(x-(2\I+c)t\big),
\\ \label{3.12} &
h_3=-g(2\I-c+\mu), \quad h_4=-(2\I-c-\mu).
\end{align}
Hence, relations \eqref{2.17} and \eqref{2.18} yield
\begin{align}\nn
\Lam_1(x,t)^*\Lam_2(x,t)=&-\exp\{\I(2cx-2c^2t+t)\}
\\ & \nn
\times\big(|g|^2(2\I-c+\mu)\exp\{\I(\mu-\ov{\mu})(x-ct)+2(\mu+\ov{\mu})t\}
\\ & \nn
+(2\I-c-\mu)\exp\{-\I(\mu-\ov{\mu})(x-ct)-2(\mu+\ov{\mu})t\}
\\ & \nn
+g(2\I-c+\mu)\exp\{\I(\mu+\ov{\mu})(x-ct)+2(\mu-\ov{\mu})t\}
\\ & \label{3.13}
+\ov{g}(2\I-c-\mu)\exp\{-\I(\mu+\ov{\mu})(x-ct)-2(\mu-\ov{\mu})t\}
\big).
\end{align}
Relations \eqref{2.9}, \eqref{2.17} and \eqref{2.18} imply that
\begin{align}\nn
S(x,t)=&\frac{1}{4}\big(|\Lam_1(x,t)|^2+|\Lam_2(x,t)|^2\big)
\\ \nn = &
\frac{1}{4}\big((|g|^2+|h_3|^2))\exp\{\I(\mu-\ov{\mu})(x-ct)+2(\mu+\ov{\mu})t\}
\\ \nn  &
+(1+|h_4|^2)\exp\{-\I(\mu-\ov{\mu})(x-ct)-2(\mu+\ov{\mu})t\}
\\ \nn  &
+(g+h_3\ov{h_4})\exp\{\I(\mu+\ov{\mu})(x-ct)+2(\mu-\ov{\mu})t\}
\\  \label{3.14} &
+(\ov{g}+h_4\ov{h_3})\exp\{-\I(\mu+\ov{\mu})(x-ct)-2(\mu-\ov{\mu})t\}\big).
\end{align}
Choosing the branch of the square root, where $\Im(\mu)<0$, we see that the first
exponents in the sums of such in \eqref{3.13} and \eqref{3.14} are the leading terms
in the case $x\to \infty$ and the second exponents are the leading terms
in the case $x\to -\infty$. Therefore, using \eqref{2.10} and setting
\begin{align}\label{3.15}&
\psi(x,t)=\exp\{-\I(2cx-2c^2t+t)\}v(x,t),
\end{align}
we see that
\begin{align}\label{3.16}&
\lim_{x\to \infty}\psi(x,t)=\I-\frac{8|g|^2(2\I-c+\mu)}{|g|^2+|h_3|^2}=\I-\frac{8(2\I-c+\mu)}{1+|2\I-c+\mu |^2},
\end{align}
and
\begin{align}\label{3.17}&
\lim_{x\to -\infty}\psi(x,t)=\I-\frac{8(2\I-c-\mu)}{1+|h_4|^2}=\I-\frac{8(2\I-c-\mu)}{1+|2\I-c-\mu |^2}.
\end{align}
\begin{Cy}\label{CyStep} For $\psi(x,t)$ of the form \eqref{3.15}, where the NLS solution $v(x,t)$ is determined
by the formulas \eqref{2.10} and \eqref{3.11}--\eqref{3.14}, the limits \eqref{3.16} and \eqref{3.17} exist
for $x$ tending to $+\infty$ and $-\infty$, respectively. These limits coincide if and only if $c=0$. When
$c\not=0$, we speak about step-like solutions.
\end{Cy}

In order to show that the right-hand sides of \eqref{3.16} and \eqref{3.17} coincide only for $c=0$, we note
that the equality there is equivalent to the equality
$$(2\I-c+\mu)\big(1-(2\I-c-\mu)(2\I+c+\ov{\mu})\big)=(2\I-c-\mu)\big(1-(2\I-c+\mu)(2\I+c-\ov{\mu}),$$
which, in view of \eqref{3.11}, is equivalent to the equality $2(\mu +\ov{\mu})=0$. Using again the
expression for $\mu$ in \eqref{3.11}, we see that the last equality is equivalent to $c=0$.
\begin{Rk} \label{RkStep} Step-like solutions are of interest and are studied in the literature although such solutions are absent in the
mentioned above works on rogue waves.
\end{Rk}

In the Examples \ref{Ee1}, \ref{Ee2} and \ref{Ee3} above, we considered the case  $
n= \nobreak  1$. In a partly similar to Example \ref{Ee1}
case, where $n=2$, $a=1, \,\, c=0, \,\,  d=1/2$,   $ A=2 \I I_2,  \,\, h_1=\begin{bmatrix}g \\ 0 \end{bmatrix}, \,\, h_2=\begin{bmatrix}1 \\ -1 \end{bmatrix}$,
$Q=\I \mu I_2, \,\, \mu =\sqrt{3}>0$, one obtains $v(x,t)=v_0(x,t)$.
Interesting GBDT transformations $v(x)$ of the constant seed, where $n=2$ and $A$ is a $2\times 2$ Jordan cell, are treated in 
\cite[Example~4.4]{ALS1}.  The 
asymptotical behaviour of $v(x)$ is described in that case in \cite[Proposition~4.5]{ALS1}. However, that behaviour greatly differs from the
cases of asymptotics discussed in this work.

A class of the $N$-modulation solutions of NLS generated by the initial solution $v_0=\E^{\I t}$ (i.e., $c=0$)
was constructed in \cite{SaA2}. Let us construct a much wider class of such solutions for $c\not=0$.
\begin{Ee}\label{Ee2+}  Assume that $v_0$  is given by \eqref{1.2}, $n\in \BN$ and 
\begin{align}\label{3.19-}&
 A=\diag\{\cla_1,\cla_2, \ldots, \cla_n\}, \quad \cla_k=c+\I b_k \quad (b_k \in \BR, \,\,1\leq k\leq n).
\end{align} 
According to \eqref{2.11},
$Q=\diag\{\clq_1,\clq_2,\ldots,\clq_n\}$, where $\clq_k^2=|a|^2-b_k^2$  $\break (1\leq k\leq n)$. We will assume that $\clq_k$
are commensurable with $c$:
\begin{align}\label{3.18}&
b_k^2=|a|^2-\frac{r_k^2}{\ell_k^2}c^2>0 \quad (r_k, \ell_k \in \BN), \quad \clq_k=\pm \frac{r_k}{\ell_k}c.
\end{align}
Moreover, we assume that  \eqref{2.27--} holds, where $\cll_k$ are the rows of $\Lam$.
\end{Ee}
It follows that 
\begin{align}\label{3.19}&
\I\clr(x,t)=\I \, \diag\{\clr_1(x,t),\ldots \clr_n(x,t)\}, \quad \clr_k(x,t)=\clq_kx-\clq_k(2c+\I b_k)t.
\end{align}
In view of \eqref{2.17}, \eqref{2.18} and \eqref{3.19}, $v_0(x,t)$, $\Lam_1(x,t)$ and $\Lam_2(x,t)$ are periodic with respect to $x$
with the period 
\begin{align}\label{3.20}&
T=(2\pi /c)\prod_{k=1}^n\ell_k.
\end{align}
According to  Corollary \ref{CyDiag}, $S(x,t)$ has the same period and is invertible. Thus, $S(x,t)^{-1}$ has the same
period $T$ given by \eqref{3.20}. Finally, it follows from \eqref{2.10} that the corresponding GBDT- transformed solutions
$v(x,t)$ are $N$-modulation solutions with the period $T$.
\section{Baker-Akhiezer functions and evolution of the Weyl-Titchmarsh functions} \label{Evol}
\setcounter{equation}{0}
{\bf 1.} According to \eqref{2.10--}, we need to know $w_0(x,t,z)$ satisfying \eqref{2.10!} in order to construct Baker-Akhiezer functions $w(x,t,z)$ 
satisfying \eqref{2.10-}, that is,  Baker-Akhiezer functions corresponding to the 
GBDT-generated solutions of NLS. Recall that $G_0$ and $F_0$ in \eqref{2.10!} are given by \eqref{2.2} and \eqref{2.3}, where one sets
$v(x,t)=v_0(x,t)$ and $v_0$ is given by \eqref{1.2}.
\begin{Pn}\label{Pnw0}  The matrix function $w_0$ of the form
\begin{align}\label{4.1}&
w_0(x,t,z):=\E^{\I (c x+dt) j}Z(z)\exp\big\{\I  \zeta(z)\big(x-(z+c)t\big) j\big\}, \quad {\mathrm{where}}
\\ & \label{4.2}
\zeta(z)=\sqrt{(z-c)^2+|a|^2}, \quad Z(z)=\begin{bmatrix}\I a & \I a \\
z-c-\zeta(z) & z-c+\zeta(z)
\end{bmatrix},
\end{align}
 satisfies systems \eqref{2.10!} $($i.e., $w_0$ is the initial Baker-Akhiezer function$)$.
\end{Pn}
{\it Proof}. In view of \eqref{4.1}, the equation $\frac{\p}{\p x}w_0(x,t,z)=G_0(x,t,z)w_0(x,t,z)$ is equivalent to the equality
\begin{align}\label{4.3}&
\I \E^{\I (c x+dt) j}(cjZ(z)+\zeta(z)Z(z)j)=(\I z j+ j V_0(x,t))\E^{\I (c x+dt) j}Z(z),
\end{align}
which is, in turn, equivalent to
\begin{align}\label{4.3+}&
\I\big(z j Z(z)-cj Z(z)-\zeta(z)Z(z)j\big)+\E^{-\I (c x+dt) j} j V_0(x,t)\E^{\I (c x+dt) j}Z(z)=0.
\end{align}
Taking into account \eqref{4.2}, we see that
\begin{align}\label{4.4}&
(z-c-\zeta(z))(z-c+\zeta(z))=-|a|^2.
\end{align}
Using \eqref{4.2} and \eqref{4.4}, we obtain
\begin{align}\label{4.5}&
z j Z(z)-cj Z(z)-\zeta(z)Z(z)j=\begin{bmatrix}\I a(z-c-\zeta(z)) & \I a (z-c+\zeta(z))\\
|a|^2 & |a|^2
\end{bmatrix}.
\end{align}
It follows from \eqref{1.2} that 
\begin{align}\label{4.6}
\E^{-\I (c x+dt) j} j V_0(x,t)\E^{\I (c x+dt) j}Z(z)&=\begin{bmatrix}0 & a \\-\ov{a} &0
\end{bmatrix}Z(z)
\\ & \nn
=-\I \begin{bmatrix}\I a(z-c-\zeta(z)) & \I a (z-c+\zeta(z))\\
|a|^2 & |a|^2
\end{bmatrix}.
\end{align}
Formula \eqref{4.3+} is immediate from \eqref{4.5} and \eqref{4.6}, and so the first equation in \eqref{2.10!}
is proved.

For $w_0$ given by \eqref{4.1}, let us prove the second equation in \eqref{2.10!}.  Formulas \eqref{2.3}, \eqref{2.12} and \eqref{1.2}
yield
\begin{align}\nn
F_0(x,t,z)&=\I \left(-z^2+\frac{|a|^2}{2}\right)j-(z+c )jV_0(x,t)
\\ \label{4.7}&
=-\big(\I(z^2-c^2-d)j+(z+c)jV_0(x,t)\big).
\end{align}
Hence, the second equation in \eqref{2.10!} is equivalent to the equality
\begin{align}\nn &
\big(\I(z^2-c^2-d)j+(z+c)jV_0(x,t)\big)\E^{\I (c x+dt) j}Z(z)
\\ \label{4.8}&
=-\I \E^{\I (c x+dt) j}(d jZ(z)-(z+c)\zeta(z)Z(z)j).
\end{align}
The equality \eqref{4.8} is equivalent to the equality
\begin{align}\label{4.9} &
\I(z-c)jZ(z)-\I \zeta(z)Z(z)j+\E^{-\I (c x+dt) j}jV_0(x,t)\E^{\I (c x+dt) j}=0.
\end{align}
The equality \eqref{4.9} easily follows from \eqref{4.5} and \eqref{4.6}. Hence,  the second equation in \eqref{2.10!}
is proved. $\Box$
\begin{Rk} Now, Baker-Akhiezer function corresponding to the triple $\break \{A, S(0,0), \Lam(0,0)\}$ is given by the relation \eqref{2.10--}
as well as by the formulas \eqref{2.9+}, \eqref{4.1}, where $\Lam(x,t)$ and $S(x,t)$ in \eqref{2.9+} are given by \eqref{2.17}, \eqref{2.18} and \eqref{2.7},
\eqref{2.8}, respectively.
\end{Rk}

{\bf 2.} In order to define Weyl functions of the skew-selfadjoint auxiliary Dirac systems \eqref{1.3} (see \cite{ALS1, SaSaR} and the references therein)
we need the notation $\BC_M:=\{z: \, z\in \BC, \,\, \Im(z)>M>0\}.$ Here,  we consider the case $p=1$ and define Weyl functions for this case although the generalization
for $p>1$ is evident. By $W(x,z)$, we denote the fundamental solution  of the Dirac system $W^{\prime}(x,z)=(\I z j+j V(x))W(x,z)$ normalised by the equality
$W(0,z)=I_2$. Recall that  $j=\begin{bmatrix}1 & 0 \\   0 & - 1\end{bmatrix}$ and $V(x)=\begin{bmatrix}0 & v(x) \\  \ov{v(x} & 0\end{bmatrix}$.
\begin{Dn} \label{DnW2}  A  function $\vp(z)$, which is  holomorphic in $\BC_M$
$($for some  $M>0)$ and satisfies the inequality
\begin{align}&      \label{4.10}
\int_0^{\infty}
\begin{bmatrix}
1 & \vp(z)^*
\end{bmatrix}
W(x,z)^*W(x,z)
\begin{bmatrix}
1 \\ \vp(z)
\end{bmatrix}dx< \infty , \quad z\in \BC_M,
\end{align}
is called a Weyl function of  the skew-selfadjoint Dirac system on $[0,\infty)$.
\end{Dn}
For the solutions of the general type inverse problems to recover $v(x)$ from the  Weyl function see \cite {SaSaR} and references therein.

Evolution $\vp(t,z)$ of the Weyl function for the case of GBDT-transformed $v(x,t)$ instead of $v(x)$ is easily expressed (in terms
of $\Lam(0,t)$ and $S(0,t)$ given by \eqref{2.6} and \eqref{2.8}) using  \cite[Theorem 5.7]{ALS1} on the Weyl function for the case of GBDT-transformed $v(x)$.
Recall that $\zeta(z)=\sqrt{(z-c)^2+|a|^2}$. Here, we choose the branch of the square root so that $\zeta(z)\in \BC_+$ for $z\in \BC_+$.
\begin{Tm}\label{TmEvol} Let $v(x,t)$ be generated by the triple $\{A,S(0,0), \Lam(0,0)\}$ and initial seed $v_0(x,t)$ of the form \eqref{1.2}.
Then, the Weyl function $\vp(t,z)$ of the corresponding skew-selfadjoint Dirac system on $[0,\infty)$ is given by the formula
\begin{align}\label{4.11} &
\vp(t,z)={\E^{-2\I dt}}\Big(D(z)+\big(D(z)C_1-C_2\big)(A^{\times}-z I_n)^{-1}B(z)\Big)\big/(\I a),
\end{align}
where $D(z)=z-c-\zeta(z)$,
\begin{align}\label{4.12} &
 C_1=\frac{1}{a}{\E^{-2\I dt}}\Lam_1(0,t)^*S(0,t)^{-1}, \quad C_2=\I  \Lam_2(0,t)^*S(0,t)^{-1},  
\\ \label{4.13} &
A^{\times}=A-B(z)C_1, \quad B(z)=\I a \E^{2\I dt}\Lam_1(0,t)+D(z)\Lam_2(0,t).
\end{align}
\end{Tm}
Note that the representation \eqref{4.11} of $\vp(t,z)$ is close to the {\it realisations} in \cite{GKS2} (it would be a realisation if $D$ and $B$ did not depend on $z$).

\vspace{0.3em}

\noindent {\bf Data availability} Not applicable.

\vspace{0.3em}

\noindent {\bf Declarations}

\vspace{0.3em}

\noindent {\bf Conflict of interest} The author has no Conflict of interest to declare that are
relevant to the content of this article.

\end{document}